\RequirePackage{lineno}
\documentclass[twocolumn,aps,prl,showpacs,superscriptaddress,floatfix]{revtex4-1}

\usepackage{lineno}

\usepackage{graphicx}

\makeatletter
\renewcommand{\p@subsection}{}
\renewcommand{\p@subsubsection}{}
\makeatother

\begin{document}

\newcommand{\Nch}{N_{\rm ch}}
\newcommand{\dg}{\Delta\gamma}
\newcommand{\R}{R_{\Psi_{2}}}
\newcommand{\dS}{\Delta S}
\newcommand{\sigR}{\sigma_{\R}}
\newcommand{\mean}[1]{\langle#1\rangle}

\title{Comment on ``Scaling properties of background- and chiral-magnetically-driven charge separation in heavy ion collisions at $\sqrt{s}_{\rm NN}=200$ GeV (arXiv:2203.10029)''}


\author{Fuqiang Wang}
\email{fqwang@purdue.edu}
\address{Department of Physics and Astronomy, Purdue University, West Lafayette, IN 47907, USA}

\date{\today} 


\begin{abstract}
  In a recent preprint arXiv:2203.10029, the authors argue for a multiplicity 
  scaling of the squared inverse width 
  of the $\R$ distribution to claim a positive signal for the chiral magnetic effect from STAR's isobar data, in stark contradiction to the conclusion reached by the STAR Collaboration. This Comment points out the fallacy of their arguments and reinforces the STAR conclusion.
\end{abstract}

\maketitle


In their recent publication of the isobar data~\cite{STAR:2021mii}, the STAR Collaboration reported the ratios between Ru+Ru and Zr+Zr collisions of the observables $\dg/v_2$ (the azimuthal correlator $\dg$~\cite{Voloshin:2004vk} divided by the elliptic flow anisotropy $v_2$) and $1/\sigR^2$ (the squared inverse width of the $\R(\dS)$ distribution~\cite{Magdy:2017yje}). Both are lower than unity, leading the STAR Collaboration to conclude that the chiral magnetic effect (CME) is not observed in their blind analysis of the isobar data~\cite{STAR:2021mii}. In the post-blind analysis of the STAR paper~\cite{STAR:2021mii}, it was pointed out that ``[u]nder the approximation that background to $\dg$ is caused by flowing clusters with the properties of the clusters staying the same and the number of clusters scaling with multiplicity ($\Nch$), the value of $\dg$ scales with the inverse of multiplicity...it may be considered that the proper baseline for the ratio of $\dg/v_2$ between the two isobars is the ratio of the inverse multiplicities of the two systems.'' The measured isobar ratios of $\dg/v_2$ are all above this baseline by an approximately 1\%. Recent studies by STAR~\cite{Tribedy,Feng} have shown, however, that deviations from the assumption about the cluster properties and contamination of nonflow correlations can move the baseline up by approximately 1\%, leading to the conclusion that the isobar ratios of the $\dg/v_2$ measurements are consistent with background expectations. For $1/\sigR^2$, on the other hand, it was pointed out in the post-blind analysis of the STAR paper~\cite{STAR:2021mii} that {\em ``[t]he scaling relations extracted in Ref.~\cite{Choudhury:2021jwd} indicate an approximate relation...$1/\sigR^2\approx\Nch\dg$; an estimate based on the measurements from this analysis indicates this ratio for Ru+Ru over Zr+Zr to be approximately 1.02.''} In other words, the baseline for the isobar ratio of $1/\sigR^2$ is 1.02.

In a recent preprint arXiv:2203.10029 by Lacey et al., titled {\em ``Scaling properties of background- and chiral-magnetically-driven charge separation in heavy ion collisions at $\sqrt{s}_{\rm NN}=200$ GeV''}~\cite{Lacey:2022baf}, the authors argue that the $1/\sigR^2$ of the $\R(\dS)$ distribution linearly depends on the inverse multiplicity ($1/\Nch$) in the Anomalous Viscous Fluid Dynamics (AVFD) model when CME is absent, and thus the $1/\sigR^2$ should be scaled by $\Nch$ for a proper background cancellation between the isobar systems. Because the multiplicity in Ru+Ru is larger than in Zr+Zr by approximately 4.4\% in the 20-50\% centrality range~\cite{STAR:2021mii}, the multiplicity scaling would lift the isobar ratio of $\Nch/\sigR^2$ above unity, constituting a positive CME signal as the authors claim: {\em``Corrections to recent $\R(\dS)$ measurements~\cite{STAR:2021mii} that account for the background difference in Ru+Ru and Zr+Zr collisions, indicate a charge separation difference between the isobars compatible with the CME.''} This utterly contradicts the conclusion reached by the STAR Collaboration in the isobar paper~\cite{STAR:2021mii}, of which the two leading authors of the preprint arXiv:2203.10029 (Lacey and Magdy) are coauthors. In this Comment, I point out the fallacy of their arguments in the preprint.

\begin{figure}
  \includegraphics[width=0.8\linewidth]{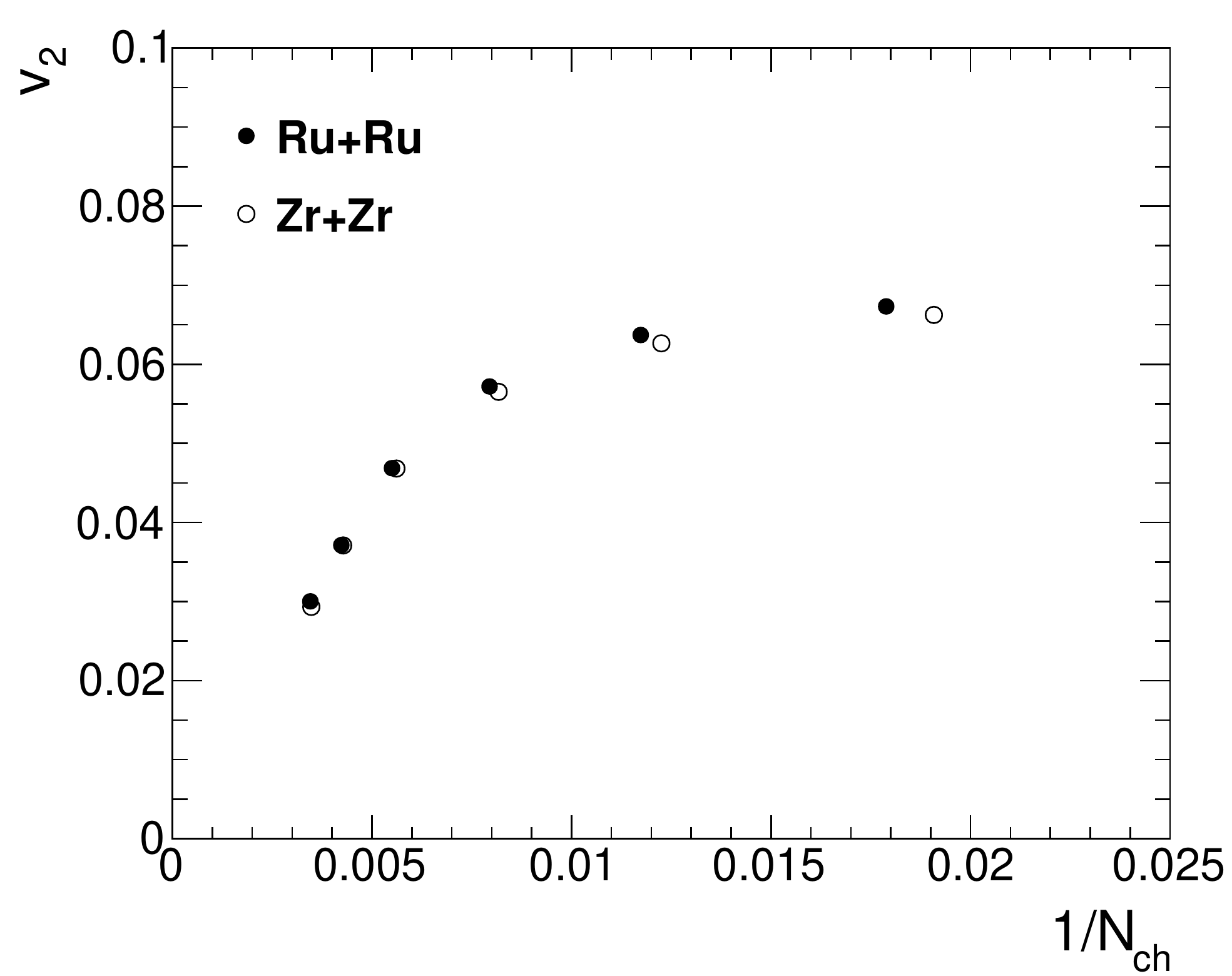}
  \caption{The average elliptic flow anisotropy $v_2$ versus the inverse multiplicity $1/\Nch$ in Ru+Ru (solid circles) and Zr+Zr (open circles) collisions in the relevant centrality range of 0-50\%. Data from Ref.~\cite{STAR:2021mii}.}
  \label{fig}
\end{figure}

First, the $1/\sigR^2$ versus $1/\Nch$ shown in Fig.~3 of the preprint~\cite{Lacey:2022baf} is apparently linear, but not proportional. This renders already the invalidity of the $\Nch$ scaling by the authors. But more importantly, the $1/\sigR^2$ does not explicitly depend on $\Nch$ as demonstrated in Refs.~\cite{Feng:2020cgf,Choudhury:2021jwd}. 
This can easily be seen as follows. The $\R$ distribution~\cite{Magdy:2017yje} is a double ratio involving four probability distributions, one in $\dS=\mean{\sin(\phi_{+}-\Psi_2)}-\mean{\sin(\phi_{-}-\Psi_2)}$, the mean sine difference of particle azimuthal angles ($\phi$) relative to the second order harmonic plane ($\Psi_2$) between positive and negative particles, another in $\dS_\perp$ which is analogous to $\dS$ but with the $\phi$ angles rotated by $\pi/2$, and the other two in $\dS$ and $\dS_\perp$, respectively, after the particle charges are randomly shuffled. The $1/\sigR^2$ is thus a combination of the variances of $\dS$ and $\dS_\perp$, effectively equal to the $\mean{\cos(\phi_\alpha-\Psi_2)\cos(\phi_\beta-\Psi_2)-\sin(\phi_\alpha-\Psi_2)\sin(\phi_\beta-\Psi_2)}$ between opposite-sign and same-sign charge pairs of $\alpha,\beta$ particles~\cite{Voloshin,Choudhury:2021jwd,Feng:2020cgf}, which is equivalent to the $\dg$ variable~\cite{Voloshin:2004vk}. 
Additionally, the $\dS$ and $\dS_\perp$ are both scaled by the width of the charge-shuffled distribution in $\dS$ to minimize particle number fluctuations~\cite{Magdy:2017yje}. This introduces a multiplicative factor $(\sqrt{\Nch})^2$, leading to the following algebraic relation $1/\sigR^2\approx\Nch\dg$~\cite{Choudhury:2021jwd,Feng:2020cgf}. This relation is explicitly stated in the STAR isobar paper~\cite{STAR:2021mii} (quoted in the first paragraph of this Comment). Because $\dg$ is approximately inversely proportional to $\Nch$, the $1/\sigR^2$ on first order does not have an explicit $\Nch$ dependence. On the other hand, $\dg$ is proportional to $v_2$~\cite{Voloshin:2004vk,Wang:2009kd}, so the $1/\sigR^2$ explicitly depends on $v_2$. Because of this $v_2$ dependence and because $v_2$ varies with multiplicity (see Fig.~\ref{fig}), a spurious $1/\Nch$ dependency of $1/\sigR^2$ results.

Figure~\ref{fig} shows $v_2$ versus $1/\Nch$ in various centralities in the range of 0-50\% of isobar collisions, taken from Ref.~\cite{STAR:2021mii}. In heavy ion collisions, quantities (such as $\Nch$, $v_2$, and $\sigR$) vary with centrality and because of that, those quantities correlate with each other one way or the other. In a given collision system, one can choose whatever variable to describe the trends of their data; however, in comparing two different systems, one has to be careful.
The strength of the isobar collisions in searching for the CME relies on comparison between the two systems and cancellation of their backgrounds. Because the backgrounds are slightly different between the isobar systems~\cite{Xu:2017zcn,Li:2018oec,Xu:2021vpn,STAR:2021mii}, some sort of scaling is needed to remove this difference. How the background should be scaled depends critically on the physics driving the background.
As demonstrated in Refs.~\cite{Feng:2020cgf,Choudhury:2021jwd} and recapitulated in the preceding paragraph, the $v_2$ is the driving physics underlying the $1/\sigR^2$ background; the $1/\sigR^2$ isobar ratio should be scaled down by the $v_2$ ratio between Ru+Ru and Zr+Zr collisions.
This will move the isobar ratio of $1/\sigR^2$ down by an approximately $-2$\%, further below unity. This is explicitly stated, in effect, in the STAR isobar paper~\cite{STAR:2021mii} as quoted in the first paragraph of this Comment. This reinforces the STAR conclusion that no CME is observed in the $\R$ observable. In fact, it puts even more stringent constraint on what can be learned from the $\R$ observable.
On the other hand, as clearly shown in Fig.~\ref{fig}, the changes from one isobar system to the other in $v_2$ and in $1/\Nch$ are in opposite directions.
If one is not discriminative, taking the spurious correlation between $1/\sigR^2$ and $1/\Nch$ as evidence of a casual relationship and applying $\Nch$ scaling to the isobar ratio of $1/\sigR^2$, as the authors did, then an exactly opposite conclusion is reached. 

The authors themselves proposed the normalization of $\dS$ and $\dS_\perp$ by the Gaussian width of the charge-shuffled distribution~\cite{Magdy:2017yje}. So it is clear to them that the $\R$ distribution is a double ratio of four distributions, {\em all} with approximately unity Gaussian width, {\em irrespective} of the $\Nch$. It is not readily expected to have an explicit $\Nch$ dependence; indeed, algebra~\cite{Choudhury:2021jwd,Feng:2020cgf} shows no explicit $\Nch$ dependence in $1/\sigR^2$. 
On the other hand, the $\R$ variable is designed to compare in-plane and out-of-plane distributions, so it {\em must} depend on the average $v_2$ of the events. This was amply demonstrated in the literature~\cite{Feng:2018chm,Feng:2020cgf,Choudhury:2021jwd}. The authors performed event-shape-engineering (ESE) analysis using the $q_2$ vector~\cite{Schukraft:2012ah} and used the apparent $q_2$ independence to argue that $1/\sigR^2$ is insensitive to $v_2$. In fact, the ESE results in the STAR isobar paper~\cite{STAR:2021mii} indicate an decreasing trend of $1/\sigR^2$ with increasing $q_2$ whereas $v_2$ increases with $q_2$. This is rather peculiar and the STAR paper warns~\cite{STAR:2021mii} that {\em``[f]urther studies may be needed to understand the physics behind the observed behavior of the widths of $\R$ on $q_2$''}. As there should be no question that any variable exploiting in-plane and out-of-plane difference should depend on the event $v_2$, and it is known that ESE results may be difficult to interpret with limited acceptance because of short-range correlations, as is the case for STAR, the authors' keeping promoting the $v_2$ independence narrative of their observable is a mystery. 

In conclusion, the multiplicity $\Nch$ is already factored out by the normalization of the $\R$ distribution to the charge-shuffled width~\cite{Magdy:2017yje}. Scaling the squared inverse width $1/\sigR^2$ by $\Nch$ in arXiv:2203.10029~\cite{Lacey:2022baf} is spurious and the conclusion from it is bogus. The correct background scaling factor is the elliptic flow anisotropy $v_2$, and no CME is observed in the isobar ratio of $1/\sigR^2$ or its $v_2$ scaled one, as stated in the STAR isobar paper~\cite{STAR:2021mii}. 

This work was supported by the U.S.~Department of Energy under Grant No.~DE-SC0012910.

\bibliographystyle{unsrt} 
\bibliography{ref}

\end{document}